\documentclass[10pt,twocolumn,
amssymb, nobibnotes, notitlepage,nofootinbib aps, prd]{revtex4}

\usepackage{amsmath}
\usepackage{amssymb}
\usepackage{amsthm}
\usepackage[pdftex]{color}
\usepackage[pdftex,colorlinks,citecolor=blue,linkcolor=blue,urlcolor=blue]{hyperref} 
\usepackage{graphicx}
\usepackage{dcolumn} 
\usepackage{bm} 

\usepackage{longtable}

\usepackage{ulem}   
\usepackage{comment} 
\usepackage{datetime}

\usepackage{tabularx}
\usepackage{float}
\restylefloat{table}

\newcommand{\beq}{\begin{equation}}
\newcommand{\eeq}{\end{equation}}
\newcommand{\bea}{\begin{eqnarray}}
\newcommand{\eea}{\end{eqnarray}}
\newcommand{\barr}{\begin{array}}
\newcommand{\earr}{\end{array}}

\long\def\begincomment#1\endcomment{}

\newcommand{\g}{\gamma}
\newcommand{\p}{\psi}

\newcommand{\tr}{\mathop{\mathrm{tr}}}

\newtheorem{theorem}{Theorem}
\newtheorem{proposition}{Proposition}

\begin{document}

%\sffamily

\title{ Pair production of  Dirac particles in a   $d+1$-dimensional  noncommutative space-time
}

\author{Dine Ousmane Samary} \email{dsamary@perimeterinstitute.ca}  
%\footnote{(Email:dsamary@perimeterinstitute.ca)}  
\affiliation{Perimeter Institute for Theoretical Physics, Waterloo, ON, N2L 2Y5, Canada}
\affiliation{International Chair in Mathematical Physics and Applications (ICMPA-UNESCO Chair), University of Abomey-Calavi,
072B.P.50, Cotonou, Republic of Benin}

\author{Emanonfi Elias N'Dolo} 
\email{emanonfieliasndolo@yahoo.fr}
\affiliation{International Chair in Mathematical Physics and Applications (ICMPA-UNESCO Chair), University of Abomey-Calavi,
072B.P.50, Cotonou, Republic of Benin}

\author{Mahouton Norbert Hounkonnou}
\email{norbert.hounkonnou@cipma.uac.bj}
\affiliation{International Chair in Mathematical Physics and Applications (ICMPA-UNESCO Chair), University of Abomey-Calavi,
072B.P.50, Cotonou, Republic of Benin}

\date{\today}
%\date{\currenttime, \today}

\begin{abstract}
This work addresses the  computation of the propability of  fermionic particle pair production   in $(d+1)-$ dimensional noncommutative Moyal space. Using the Seiberg-Witten maps that establish  relations between noncommutative and commutative field variables,  up to the
 first order in the noncommutative parameter $\theta$, we derive the probability density of  vacuum-vacuum pair production of  Dirac particles. The cases of constant electromagnetic,  alternating time-dependent and space-dependent electric fields are considered and discussed.
\end{abstract}


\maketitle
%\tableofcontents

\section{Introduction}
Noncommutative  field theory (NCFT), arising  from  noncommutative (NC) geometry, has been the subject of intense studies, owing to its importance in the description of quantum gravity phenomena.  More precisely, the concepts of noncommutativity in fundamental physics have deep motivations which originated from
the fundamental properties of the Snyder space-time \cite{Snyder:1946qz}.  Further, the results by Connes, Woronowicz and Drinfel’d \cite{Connes:1994yd,Connes:1990qp,Woronowicz:1987wr}  provided a clear definition of  NC geometry, thus bringing a new stimulus in this area. The
 NC geometry arises as a possible scenario for the short-distance behaviour of physical theories (i. e. the Planck length scale $\lambda_p=\sqrt{\frac{G\hbar}{c^3}}\approx 1,6\cdot 10^{-35}\mbox{ meters}$), see \cite{Doplicher:1994tu,Majid:1999tc,Szabo:2001kg} and references therein. This fundamental unit of length marks the scale of energies and distances at which the non-locality
of interactions has to appear and a notion of continuous space-time becomes meaningless \cite{Doplicher:1994tu,Majid:1999tc,Banburski:2013jfa}. 
One of the important implications of  noncommutativity is the Lorentz violation symmetry   in more than two dimensional
space-time \cite{Ferrari:2005ng,Imai:2000kq,Carmona:2002iv}, which, in part, modifies the dispersion relations \cite{Jackiw:2001dj}.
It led to new developments in quantum electrodynamics (QED) and Yang-Mills (YM) theories in the
NC variable function versions \cite{Raasakka:2010ev,Liao:2002kd}. The same observation appears in the framework of  string theory \cite{Seiberg:1999vs,Bozkaya:2002at}.  Also,   the quantum Hall effect well illustrates  the NC quantum mechanics of space-time \cite{Harms:2006dv,Scholtz:2005vg} (and references therein).

In this work, we  use a NC star product  obtained by replacing the ordinary product of functions by
 the 
%NC  star product also called 
Moyal star product as follows:
\bea\label{star}
&&f\star g= {\rm\bf m}\Big[\exp\Big(\frac{i}{2}\theta^{\mu\nu}\,\partial_\mu\otimes\partial_\nu\Big)(f\otimes g)\Big],
\eea
 where $f$, $g\in C^{\infty}(\mathbb{R}^D)$,  ${\rm\bf m}(f\otimes g)=f\cdot g;$ 
$\theta^{\mu\nu}$ stands for a skew-symmetric tensor characterizing the NC behaviour of the space-time, and has the Planck's length square dimension, i.e.  $[\theta]\equiv [\lambda_p^2].$ 
%The space-time dimension is $D=d+1$.
The star product   \eqref{star} 
%of two functions $f$ and $g$ 
satisfies the useful integral relation
\bea
\int\,d^Dx\, (f\star g)(x)&=&\int\,d^Dx\, (g\star f)(x)\cr
&=&\int\,d^Dx\, f(x)\, g(x).
\eea
It provides the following commutation relation between the  coordinate functions:
\bea
[x^\mu,x^\nu]_\star=x^\mu\star x^\nu-x^\nu\star x^\mu=i\theta^{\mu\nu},\quad  x^\mu\in\mathbb{R}^D.
\eea
 For convenience, we  write the tensor $(\theta^{\mu\nu})$ in the following form:
\bea\label{matrix1}
(\theta^{\mu\nu})=\left(\begin{array}{ccccc}
0&\theta&\cdots&0&0\\
-\theta&0&\cdots&0&0\\
\vdots&\vdots&&\vdots&\vdots\\
0&0&\cdots&0&\theta\\
0&0&\cdots&-\theta&0
\end{array}\right),\quad \theta\geq 0.
\eea
The relation \eqref{matrix1} means that the time does not commute with NC spatial coordinates and  the dimension $D$ of the NC space-time is even. 
Recall that two main problems arise when one tries to implement the electromagnetism in a
NC geometry: the loss of causality due to the appearance of derivative couplings in
the Lagrangian density and, more fundamentally, the violation of Lorentz invariance exhibited by plane
wave solutions \cite{Jackiw:2001dj,Berrino:2002ss}. 

Like in ordinary quantum mechanics, the NC coordinates  satisfy  the coordinate-coordinate version of the Heisenberg  uncertainty relation, namely $\Delta x^\mu\Delta x^\nu\geq \theta$, and then make the space-time a quantum space.  This idea leads to the concept of quantum gravity, since quantizing space-time leads to quantizing gravity.   Apart from the overall results about QED and YM theory in NC space-time, it turns out to be important to understand how noncommutativity modifies the probability of pair production of fermionic particles. This is the task we shall deal with  in this work. 

A pair production refers to  the creation of an elementary particle and its antiparticle, usually when a  neutral boson interacts with a nucleus or another boson. Nevertheless a static electric field in an empty space can create electron-positron  pairs. This effect, called the Schwinger effect \cite{Schwinger:1948iu}, is currently  on the verge of being experimentally
verified.  Recently, the vacuum-vacuum transition amplitude and its probability density were computed in four, three and two dimensional space-time within  constant and alternating electromagnetic (EM) fields (\cite{Gavrilov:1996pz}-\cite{Hounkonnou:2000im} and reference therein).
The related questions have been discussed and  gained considerable attention in the researchers community.

In this work, we provide   the NC version of pair production of  Dirac particles. Specially, we derive the exact expression for the probability density of particle production by an external field.  This  establishes  a relation with  important analytical results which were previously obtained in the ordinary space-time, spread in the literature  \cite{Schwinger:1948iu, Gavrilov:1996pz,Lin:1998rn,
Lin:1999bb,Lin:2000rh,Hounkonnou:2000im}. In particular, the case of $(d+1)$-dimensional space-time which have been derived in \cite{Gavrilov:1996pz} and \cite{Hounkonnou:2000im} are extended in noncommutative case.

The paper is organized as follows.  In   section \eqref{sec1}, we  quickly review the  Seiberg-Witten maps giving a relation between NC field variables and  commutative ones \cite{Seiberg:1999vs,Fidanza:2001qm,Ulker:2012yk}.  Here we also expose the main result about  gauge theory in NC space, that allows us to write the NC Lagrangian density of the  Dirac particle (oupling to EM field) with the commutative field variables. In section \eqref{sec3} we compute the probability density of pair production of a
Dirac particle in  constant EM fields.  In Section \eqref{secnew} we give the  discussions and the comments of our result. This section also contains a similar analysis  in the case of an alternating (EM) field. In 
Section \eqref{sec4}, we conclude and made some useful remarks.  Appendices \eqref{Appendix1} and \eqref{Appendix2} are enriched by the proofs of 
%two Appendices in which we offer proof of the 
key theorems set in the main part of this paper.

\section{NC gauge theory and Seyberg-Witten maps}\label{sec1}
Like in an ordinary space-time, a gauge theory can be defined on a NC space-time \cite{Madore:2000en}. In the sequel,  the NC variables are denoted with a ``hat'' notation. Let $\mathcal A_\theta$ be a Moyal algebra of functions and $\hat X\in\mathcal A_\theta$ be the covariant coordinate expressed in terms of gauge potential $\hat A\in\mathcal A_\theta$ as:
\bea
\hat X=\hat x+\hat A.
\eea
For an arbitrary function $\hat\psi\in\mathcal A_\theta$, the infinitesimal gauge transformation with parameter $\hat\Lambda\in\mathcal A_\theta$ is $\hat\delta\hat\psi=i\hat\Lambda\star\hat\psi$. The  infinitesimal variation of the gauge potential can be written  as
\bea
\hat\delta_{\hat\Lambda} \hat A^\mu=i[\hat\Lambda, \hat A^\mu]_\star-i [\hat x^\mu,\hat\Lambda]_\star.
\eea
Also the NC  Faraday tensor is given by
\bea
\hat F_{\mu\nu}=\partial_\mu \hat A_\nu-\partial_\nu \hat A_\mu-i[\hat A_\mu, \hat A_\nu]_\star.
\eea
Its infinitesimal variation is
\bea
\hat\delta\hat F_{\mu\nu}=i[\hat\lambda,\hat F_{\mu\nu}]_\star.
\eea

Besides, the functional action for a  Dirac particle on NC space-time can be defined as follows:
\bea\label{action}
S&=&\int_{\mathbb{R}^D}\,d^Dx\,\mathcal{L}(\hat{\bar\psi},\hat\psi),\\ 
\mathcal{L}(\hat{\bar\psi},\hat\psi) &=&\hat{\bar{\psi}}\star i\g^{\mu}\hat{ D}_{\mu}\hat{\p} -m\hat{\bar{\p}}\star\hat{\p}.
\eea
In this expression $\hat{\p}$ and $\hat{\bar\psi}$ are   the Dirac spinor and its associated Hermitian conjugate, respectively. The $\gamma$'s are the Dirac matrices which satisfy the Clifford algebra: $\{\gamma^\mu,\gamma^\nu\}=2\eta^{\mu\nu}$, and are  given explicitly in terms of Pauli matrices $\sigma^i,\,i=1,2,3,$ by:
\bea
\g^0=\left(\begin{array}{cc}
1_2&0\\
0&-1_2
\end{array}\right),\,
\g^i=\left(\begin{array}{cc}
0&\sigma^i\\
-\sigma^i&0
\end{array}\right).
\eea
  The covariant derivative $\hat{ D}_{\mu}$ is expressed as:
\beq
\hat{ D}_{\mu}=\partial_\mu-i\hat A_\mu\star.
\eeq
We choose $\hbar=c=1$ and  take the charge of particle equal to
the unit value,  i.e. $q_e=1$.
The Lagrangian $\mathcal{L}(\bar\psi,\psi)$ describes the propagation of the massive fermion (electron in this case) and their
interaction with photons via the covariant derivative $\hat{D}_\mu$.
In this work, we treat in detail   the case when the dimension of the space-time is equal to $D=3+1$.  The results for the cases where, $D=1+1$, and, more generally, $D=d+1,$  computed in a similar way, are given. Note that, despite the singularity exibits by the matrix $(\theta^{\mu\nu})$ in the case of odd dimensions, the probability of pair production is well defined with the same analysis.

In what follows we give  the Seiberg-Witten maps at the first order of perturbation in $\theta$ \cite{Seiberg:1999vs,Fidanza:2001qm,Ulker:2012yk}. We write  the NC field variables as function of commutative variables:
\bea\label{expand}
&&\hat{\p} = \p-\frac{1}{4}\theta^{\kappa\lambda}A_{\kappa}(\partial_{\lambda} + D_{\lambda})\p  \\
&&\label{expand1}\hat{\bar{\p}} = \bar{\p}-\frac{1}{4}\theta^{\kappa\lambda}{A}_{\kappa}(\partial_{\lambda} + {D}_{\lambda})\bar\psi
\\
&&\label{expand2}\hat{A}_{\mu} = A_{\mu}-\frac{1}{4}\theta^{\kappa\lambda}\big\{A_{\kappa},\partial_{\lambda}A_{\mu}+F_{\lambda\mu}\big\}.
\eea
By substituting the expressions \eqref{expand}, \eqref{expand1} and \eqref{expand2} in the action \eqref{action}, we get, at the first order in $\theta,$ 
% the following Lagrangian density:
\bea\label{Lagrangian}
&&\quad\quad\quad\quad \quad\mathcal{L}(\bar\psi,\psi)=\cr
&& i\gamma^\mu\Big[\bar\psi(\partial_\mu-iA_\mu)\psi+\frac{i}{2}\theta^{\alpha\beta}\partial_\alpha\bar\psi
\partial_\beta(\partial_\mu-iA_\mu)\psi\cr
&&-\frac{1}{4}\theta^{\alpha\beta}\bar\psi\partial_\mu\Big(
A_\alpha(\partial_\beta+D_\beta)\psi\Big)+\frac{1}{2}\theta^{\alpha\beta}\bar\psi\partial_\alpha A_\mu
\partial_\beta\psi\cr
&&+\frac{i}{4}\theta^{\alpha\beta}\bar\psi A_\mu A_\alpha(\partial_\beta+D_\beta)\psi
+\frac{i}{4}\theta^{\kappa\lambda}\bar\psi\big\{A_{\kappa},\partial_{\lambda}A_{\mu}\cr
&&+F_{\lambda\mu}\big\}\psi-\frac{1}{4}\theta^{\kappa\lambda}
A_\kappa(\partial_\lambda+D_\lambda)\bar\psi(\partial_\mu-iA_\mu)\psi\Big]\cr
&-&m\Big[\bar\psi\psi+\frac{i}{2}\theta^{\mu\nu}\partial_\mu\bar\psi\partial_\nu\psi-\frac{1}{4}\theta^{\mu\nu}\bar\psi A_\mu(\partial_\nu+D_\nu)(\psi)
\cr
&&-\frac{1}{4}\theta^{\mu\nu}A_\mu(\partial_\nu+D_\nu)(\bar\psi)\psi\Big]+O(\theta^2).
\eea
In the commutative limit i.e.  $\theta\rightarrow 0, $ we recover, as expected,  the Lagrangian density $\mathcal{L}_C$ of a 
 Dirac field in an ordinary space-time associated to  the functional action $\mathcal S [\psi, \bar\psi, A]:$ 
\bea
&&\mathcal S [\psi, \bar\psi, A]=\int\,d^Dx\,\mathcal{L}(\bar\psi,\psi)\cr
&&=\int\,d^Dx\,\Big(\mathcal{L}_{C}(\bar\psi,\psi)+\mathcal{B}(\theta,A,\bar\psi,\psi)\Big),
\eea
where  the quantity $\mathcal{B}(\theta,A,\bar{\p},\p)$  depending on $\theta$  is given, after some algebra, by
\bea \label{b}                                                                                                                                        
&&\mathcal{B}(\theta,A,\bar{\p},\p) = i\gamma^\mu\theta^{\kappa\lambda}\bar\psi\Big[-\frac{1}{2}(\partial_\mu
A_\kappa)\partial_\lambda
\cr
&&+\frac{1}{2}\partial_\kappa A_\mu
\partial_\lambda+\frac{i}{2} A_\mu A_\kappa\partial_\lambda
+\frac{i}{2}A_k\partial_\lambda A_\mu\cr
&&-\frac{i}{2}A_k\partial_ \mu A_\lambda+
\frac{1}{2}
(\partial_\lambda A_\kappa)\partial_\mu-\frac{i}{2}(\partial_\lambda A_k)A_\mu\Big]\psi
\cr
&&
-\frac{m\theta^{\kappa\lambda}}{2}\bar\psi(\partial_\kappa A_\lambda)\,\psi.
\eea
Now by performing the path integral over the background fields $\psi$ and $\bar\psi$,
the vacuum-vacuum transition amplitude $\mathcal Z(A)$ is afforded by the expression:
\bea
&&\quad \quad\quad\quad\quad\quad \mathcal Z(A)=\cr
 &&\mathcal N \int D\p D\bar{\p}\exp i\Big\{ \int d^{4}x\,\Big(i\gamma^\mu\bar\psi(\partial_\mu-iA_\mu)\psi\cr
&&-m\bar\psi\psi+ \mathcal{B}(\theta,A,\bar\psi,\psi)\Big) \Big\},
\eea
in which  the normalization constant $\mathcal N$ is chosen such that  $\mathcal Z(0)=1$.   Note  that
$
 \mathcal{B}(\theta,0,1,1)=0.
$ 

Let $\mathcal M:=i\gamma^\mu D_\mu-m+ \mathcal{B}(\theta,A,1,1) +i\epsilon$. Then, we get a simpler form:
\beq
\mathcal Z(A) = \exp \left[ -\tr \ln \frac{i\gamma^\mu\partial_\mu-m  +i\epsilon}{\mathcal M}\right].
\eeq
Provided with  the above quantity, we compute the probability density amplitude $|\mathcal Z(A) |^2$ for various electromagnetic fields.

\section{Transition amplitude in the case of a constant  external EM field}\label{sec3}
In this section, we consider the EM  field,  defined in $x$ direction as ${\bf B}={B\bf e_x}$ and ${\bf E}=E{\bf e_x}$, $E>0$ and $B\geq 0$.
 The position and momentum operators  $X_\mu=(X_0,X_1,X_2,X_3)=:(X_0,X,Y,Z)$ and $P_\mu=i\partial_\mu=(P_0, P_1,P_2,P_3)$ satisfy the commutation relation: 
\beq
[X_\mu,P_\mu]=i\eta_{\mu\nu}.
\eeq
 The covariant vector $V_\mu$ is expressed with the contra-variant $V^\mu$ as $V_\mu=\eta_{\mu\nu}V^\nu$, where $(\eta)=diag(1,-1,-1,-1)$.  The covariant Faraday tensor  $F_{\mu\nu}=:\partial_\mu A_\nu-\partial_\nu A_\mu$ can be expressed as:
\bea
(F_{\mu\nu})=\left(\begin{array}{cccc}
0&E_x&E_y&E_z\\
-E_x&0&-B_z&B_y\\
-E_y&B_z&0&-B_x\\
-E_z&-B_y&B_x&0
\end{array}\right),
\eea
with $ A_\mu=(-EX,0,0,BY).$
Then, $\mathcal{B}(\theta,A,1,1)$ is obtained as:
\bea\label{ko}
&&\mathcal{B}(\theta,A,1,1)=\frac{m\theta}{2}(B-E)+\frac{i\theta}{2}\gamma^\mu \Big[i(E+B)A_\mu\cr
&&-(E+B)\partial_\mu
-iA_{\mu}(EX\partial_1+BY\partial_2)-(\partial_1 A_\mu)\partial_0\cr
&&+(\partial_2 A_\mu)\partial_3+\partial_\mu(EX)\partial_1
+\partial_\mu(BY)\partial_2\Big].
\eea
Using  the charge conjugation matrix $C = i\g^{2}\g^{0}$, the identity $ C\g_{\mu}C^{-1} = -\g_{\mu}^{t}$, and  taking into account the fact that the trace of an operator is invariant under a matrix transposition lead to
%. All this consideration given:
\beq
\mathcal Z^{t}(A) =  \exp \left[ -\tr \ln \frac{iC\gamma^\mu C^{-1} \partial_\mu+m -i\epsilon}{ \mathcal M^t}\right], 
\eeq
where $\mathcal M^t=iC\gamma^\mu C^{-1} D_\mu+m- \mathcal{B}^t(\theta,A,1,1)-i\epsilon$.
The probability density is  defined by the module of $Z(A)$ as
%\begin{widetext}
\bea
|\mathcal Z(A)|^2:=\exp\Big[-\tr\ln\frac{P^2-m^2+i\epsilon}{\mathcal M\mathcal M^t}\Big],
\eea
with
\bea
\mathcal M\mathcal M^t&=&[\gamma^\mu(P_\mu+A_\mu)]^2-m^2-m^2\theta(B-E)\cr
&+&
\mathcal{B}\gamma^\mu(P_\mu+A_\mu)-\gamma^{\mu}(P_\mu+A_\mu)\mathcal{B}^t+i\epsilon.\cr&&
\eea
%\end{widetext}
The conjugate of $\mathcal{B}(\theta,A,1,1)$, denoted by  $\mathcal{B}^t(\theta,A,1,1),$ can be then written as:
\bea
&&\mathcal{B}^t(\theta,A,1,1)=\frac{m\theta}{2}(B-E)+\frac{i\theta}{2}C\gamma^\mu C^{-1} \cr
&&\times\Big[i(E+B)A_\mu
-iA_{\mu}(EX\partial_1+BY\partial_2)\cr
&&-(E+B)\partial_\mu-(\partial_1 A_\mu)\partial_0+(\partial_2 A_\mu)\partial_3
\cr
&&+\partial_\mu(EX)\partial_1+\partial_\mu(BY)\partial_2\Big].\cr
&&
\eea
At this point it would be worth  using the identity
\bea
\ln\frac{a+i\epsilon}{b+i\epsilon}=\int_0^\infty\,\frac{ds}{s}\Big[e^{is(b+i\epsilon)}-e^{is(a+i\epsilon)}\Big]
\eea
to get 
\bea\label{29}
&&\ln\frac{P^2-m^2+i\epsilon}{\mathcal M\mathcal M^t}=\int_0^\infty\,\frac{ds}{s}e^{-is(m^2-i\epsilon)}\times\cr
&&\Big[e^{is[(P+A)^2+\frac{1}{2}\sigma^{\mu\nu}F_{\mu\nu}-m^2\theta(B-E)+\mathcal{X}(\theta)]}-e^{isP^2}\Big]\cr&&
\eea
where the operator  $\mathcal{X}(\theta)$ 
%in the previous equation need to
should be  Hermitian. We use the following commutation relations
\bea\label{Rel}
[X^n,P_1]=-niX^{n-1},\quad [P_1^n,X]=niP^{n-1},
\eea
also valid when one replaces
%The relations \eqref{Rel} are well satisfied after replaced 
$X$ by $Y$ and $P_1$ by $P_2$.
For an arbitrary operator $A,$ we can define the associated  Hermitian
operator denoted by $A_H$ as 
\bea\label{Hermit}
A_H=\frac{(A+A^\dag)}{2}.
\eea
From now,   the $H$ symbol indexing
%at the bottom of 
any operator $A$, e.g. $A_H,$ refers to the   Hermitian operator associated with $A$. We then have the following:
\begin{proposition}
The  Hermitian
operator associated with  $\mathcal{X}(\theta),$ denoted $\mathcal{X}_H(\theta)$, is given by
\bea\label{xmagie}
\mathcal{X}_H(\theta)& =& \frac{\theta}{2}\Big[iEB\g^{3}\g^{2}
+iE^{2}\g^{0}\g^{1}+iB^{2}\g^{3}\g^{2}\cr 
& +&\frac{1}{2} i(\g^{0}\g^{1}+\g^3\g^2)EB+ \g^{0}\g^{1}EBYP_{2}
\cr
&+&2E^2\g^0\g^1XP_1+\g^{0}\g^{3}(E^{2}B-EB^2)XY
\cr
&-&\g^{1}\g^{3}EBYP_{1}-(4E^{3}+3BE^2)X^{2}
\cr
&+&(2B^{3}+B^2E)Y^{2}+(4E^2+5EB)XP_{0}\cr
&+&(2B^2+3EB)YP_3-2BP_{0}^{2}
+2BP_1^2\cr
&+&2EP^{2}_{2}+2EP^{2}_{3}
\Big]. 
\eea
Further,
%and it is obvious that 
\beq
\mathcal{X}_H(\theta)=\mathcal{X}^\dag_H(\theta).
\eeq
\end{proposition}
\proof
 Taking into account the fact that the trace is invariant under matrix transposition,  and using the relation \eqref{Hermit}, the operator $\mathcal{X}_H(\theta)$ takes the given form.
$\blacksquare$

Now we focus   on  the computation of the following quantity:
\bea\label{3400}
&&\mathcal O=\langle {\bf x}|e^{is[(P+A)^2+\frac{1}{2}\sigma^{\mu\nu}F_{\mu\nu}-m^2\theta(B-E)+\mathcal{X}_H(\theta)]}|{\bf x}\rangle\cr
&&=e^{\frac{1}{2}\sigma^{\mu\nu}F_{\mu\nu}-m^2\theta(B-E)}\langle{\bf x}|e^{is[(P+A)^2+\mathcal{X}_H(\theta)]}| {\bf x}\rangle,\cr&&
\eea
where  $\sigma^{\mu\nu}=\frac{i}{2}[\gamma^\mu,\gamma^\nu]$. We use the relation
\beq
\big[\gamma(P+A)\big]^2=(P+A)^2+\frac{1}{2}\sigma^{\mu\nu}F_{\mu\nu},
\eeq
and choose the 4-vectors $|{\bf x}\rangle=|x_\mu\rangle$ such that $X_\mu|{\bf x}\rangle=x_\mu|{\bf x}\rangle$.  In the momentum representation, we get a similar relation for $P_\mu|{\bf k}\rangle=k_\mu|{\bf k}\rangle$, and obtain
\beq
\langle{\bf k}|{\bf x}\rangle=\frac{1}{(2\pi)^2}e^{i\langle{\bf x}\,,\,{\bf k}\rangle},\quad \langle{\bf x}\,,\,{\bf k}\rangle=:\sum_{i=1}^4 x_ik_i.
\eeq
%For this we need to know how $(P+A)^2$ and $\mathcal{X}_H(\theta)$ are commute in itself. 
To achieve our goal, we use 
%a related combinatoric expansion that is useful  is 
the Baker-Campbell-Hausdorff formula given by
\bea
e^{t(U+V)}&=& e^{tU}~ e^{tV}e^{t^2C_2}e^{t^3C_3}e^{t^4C_4}\cdots\cr
&=&  e^{tU}~ e^{tV}\prod_{n=2}^\infty e^{t^nC_n}
\eea
where the constants $C_n$ are given by the  Zassenhaus formula \cite{Michael,Casas}: 
\bea
C_{n+1}=\frac{1}{n+1}\sum_{j=0}^{n-1}\frac{(-1)^n}{j!(n-j)!}
ad_V^j \,ad_U^{n-j}V
\eea
with
\bea
ad_UV=[U,V],\,\, ad_U^jV=[U,ad_U^{j-1}V],\,\, ad_U^0V=V.\cr&&
\eea
Explicitly, we get
\bea  
&&e^{t(U+V)}= e^{tU}~ e^{tV} ~e^{-\frac{t^2}{2} [U,V]} ~ e^{\frac{t^3}{6}(2[V,[U,V]]+ [U,[U,V]] )} \cr
&&\times~ e^{\frac{-t^4}{24}([[[U,V],U],U] + 3[[[XU,V],U],V] + 3[[[U,V],V],V]) } \cdots,\cr&&
\eea
where the exponents of higher order in t are likewise nested.
Then, 
 %let us come to our consideration: 
 take into account the first approximation of $\theta$ in the expansion of all quantities we arrive at the expression:
\bea
&&e^{is[(P+A)^2+\mathcal{X}(\theta)]}=e^{is(P+A)^2}e^{is\mathcal{X}(\theta)}e^{T(\theta)}\cr
&&=e^{is(P+A)^2}\big(1+is\mathcal{X}_H(\theta)
+T_H(\theta)+O(\theta^2)\big)\cr&&
\eea
where, for $t=is, \, U=(P+A)^2,\,$ and $ V=\mathcal{X}_H(\theta),$ we have
\bea
T_H(\theta)&=&-\frac{t^2}{2} [U,V]_H  +\frac{t^3}{6}( [U,[U,V]_H]_H )\cr
& -&\frac{t^4}{24}([[[U,V]_H,U]_H,U]_H )+\cdots .
\eea
The expectation value of the operator $e^{is[(P+A)^2+\mathcal{X}_H(\theta)]}$ is then evaluated as
\bea
&&\langle {\bf x} |e^{is[(P+A)^2+\mathcal{X}_H(\theta)]}|{\bf x} \rangle\cr
&&=\int\, d{\bf y}\,\langle {\bf x} |e^{is(P+A)^2}|{\bf y}\rangle \langle {\bf y}|\mathcal J(\theta)|{\bf x} \rangle,\cr&&
\eea
where
$
\mathcal J(\theta)=\Big(1+is\mathcal{X}_H(\theta)
+T_H(\theta)\Big).
$
Now after expanding  $U$ as
\bea
U&=&P_0^2-P_1^2-P_2^2-P_3^2-2EP_0X-2BP_3Y\cr
&+&E^2X^2-B^2Y^2,
\eea
we can easily observe that $U=U^\dag$.  As it is the  welcome,  fortunately, we get the following statement.
\begin{proposition} Let\, $U=(P+A)^2,\,$ and $ V=\mathcal{X}_H(\theta)$. The commutation relations between $U$ and $V$ are vanished, i.e.
\beq
[U,V]_H=0,\, [[[U,V]_H,U]_H,\cdots ]_H,U]_H=0
\eeq
and therefore $T_H(\theta)=0.$ 
\end{proposition}
\proof
The proof of this proposition is simply obtained by using \eqref{Hermit} and \eqref{xmagie}.
$\blacksquare$

Finally  the quantity $\mathcal O$ is reduced to
\bea
\mathcal O
&=&\mathcal O_c+\mathcal O_{nc}(\theta),\quad \mathcal O_{nc}(0)=0
\eea
where
\bea
\mathcal O_c&=&e^{\frac{is}{2}\sigma^{\mu\nu}F_{\mu\nu}}\langle {\bf x} |e^{is(P+A)^2}|{\bf x}\rangle
\eea
and
\bea
\mathcal O_{nc}(\theta)&=&e^{\frac{is}{2}\sigma^{\mu\nu}F_{\mu\nu}}\int\, d{\bf y}\,\langle {\bf x} |e^{is(P+A)^2}|{\bf y}\rangle
\cr
&\times&\langle {\bf y}|is\big[m^2\theta(E-B)+\mathcal{X}_H(\theta)\big]|{\bf x} \rangle.
\eea

We then come to the following result:
\begin{theorem}\label{th1}
Let $\theta=: \wp\,\cdot \theta_0$  where $\wp$ is a dimensionless quantity which is bounded by two  numbers $a_1,$ and $a_2$ and such that  $\theta_0<<1$. The mass dimension of $\theta_0$ is obviously $\theta_0\equiv  [M^{-2}]$. Let $M\subset\mathbb{R}^2$ be the compact subset of $\mathbb{R}^2$ in which the following integral is convergent:
\beq
\int_{M\subset\mathbb{R}^2} \frac{dt}{t_0} dz=b\equiv [M^{-2}].
\eeq
$t_0\neq 0$ is arbitrary initial time. The trace of the expectation  value $\mathcal O$ is given by:
\bea\label{C}
\tr\mathcal O=\Big(1-\wp-\sigma(\theta_0,E,B)\Big)\tr\mathcal O_c,\cr&&
\eea
where
\bea
\sigma(\theta_0,E,B)= \frac{ 16\pi^3  b\,\wp \exp\Big[is\theta_0\mathcal G_0\Big]}{\theta_0^3s^3EB}\sqrt{\frac{B}{E}f(E,B)},\cr
\eea
\bea
\tr \mathcal O_c=-\frac{1}{4\pi^2 i}EB\cosh(Es)\cot(Bs),
\eea
 $f(E,B)$ a being a positive function given by
\bea
&&f(E,B)=\Big[4B^6+76EB^5+258 E^2 B^4\cr
&&+494 E^3
B^3 +224 E^4 B^2+12E^5B\Big]^{-1}.
\eea 
\end{theorem}
\proof The proof of this theorem is given in Appendix \eqref{Appendix1}.
$\blacksquare$

Remark that the quantity $\sigma(\theta_0,E,B)$ leads to the divergence in the limit where $B=0$ and in the limit where $\theta_0=0$. This expression do not contribute to the physical solution and then the trace of $\mathcal O$  is reduced to $\tr\mathcal O=\big(1-\wp\big)\tr\mathcal O_c$ .

\begin{widetext}
\begin{theorem}\label{th2}
The vacuum-vacuum transition probability  is
$
|Z(A)|^2=\exp\Big[-\int d{ x}\,\omega_{3+1}(x)\Big]
$
where
\bea\label{intfan}
\omega_{3+1}(x)=\frac{1}{4\pi^2i}\int_{0}^\infty\, ds\,\frac{e^{ism^2}}{s}\Bigg[\Big(1-\wp\Big)EB\coth(Es)\cot(Bs)-\frac{1}{s^2}\Bigg]
\eea
whose real part, denoted by
% of $\omega(x)$, i.e. 
$\Re_e\omega(x)=\frac{\omega+\omega*}{2},$ is given by 
\bea\label{intfant1}
\Re_{e}\omega_{3+1}(x)&=&-\frac{1}{8\pi^2i}\int_{-\infty}^\infty\, ds\,\frac{e^{ism^2}}{s}\Bigg[\Big(1-\wp\Big)EB\coth(Es)\cot(Bs)-\frac{1}{s^2}\Bigg]\cr
&=&-\frac{m^4\wp}{16\pi}+\frac{EB}{4\pi^2}\big(1-\wp\big)\sum_{k=1}^\infty\frac{1}{k}\coth\Big(k\pi\frac{B}{E}\Big)\exp\Big(-\frac{k\pi m^2}{E}\Big).
\eea
\end{theorem}
\end{widetext}
\proof  The proof of this statement is given in the Appendix \eqref{Appendix2}.
$\blacksquare$

\section{Discussion of the results}\label{secnew}
In this section, we discuss  the reported results in the theorems \eqref{th1} and \eqref{th2} and provide more comment in the  framework of   $d+1$ dimensional space-time.

${\bf (1)}$\,\,
Let 
\beq
U_k=\frac{1}{k}\coth\Big(k\pi\frac{B}{E}\Big)\exp\Big(-\frac{k\pi m^2}{E}\Big),\,k\in\mathbb{N}\setminus  \{0\}.
\eeq
We get 
\bea
\lim_{k\rightarrow\infty}\Big|\frac{U_{k+1}}{U_k}\Big|=\exp\Big(-\frac{\pi m^2}{E}\Big)<1
\eea 
and conclude that, the  corresponding serie, i.e. $\sum_{k=1}^\infty U_k$ is obviously convergent. Recall that, there exist two positive constants $a_1$ and $a_2$ such that for $a_1\leq \wp\leq a_2$, $\theta_0<<1$. Then there exist the bound on $\theta$ in which the solution \eqref{intfant1} is well defined.
\bea
-\frac{m^4 a_2}{16\pi}+\frac{EB}{4\pi^2}(1-a_2)\sum_{k=1}^\infty U_k\leq\Re_e\omega_{3+1}(x)\leq\cr
-\frac{m^4 a_1}{16\pi}+\frac{EB}{4\pi^2}(1-a_1)\sum_{k=1}^\infty U_k.
\eea

{\bf (2)}\,\, Consider $\Re_e\omega_{c,3+1}(x)$ as the probability density provided with the equation \eqref{intfant1}, in the limit where $\theta\rightarrow 0$ i.e.
\beq
\lim_{\theta\rightarrow 0}\Re_e\omega_{3+1}(x)=\Re_e\omega_{c,3+1}(x).
\eeq 
This expression corresponds to the commutative limit derived by Q-G. Lin (see \cite{Lin:1998rn}) and given by:
\beq
\Re_e\omega_{c,3+1}(x)=\frac{EB}{4\pi^2}\sum_{k=1}^\infty U_k.
\eeq
 We get 
\bea
\Re_e\omega_{3+1}(x)<\Re_e\omega_{c,3+1}(x),
\eea
and we conclude that the noncommutativity increases the amplitude $|Z(A)|^2$.   This shows the importance of noncommutativity at  high energy regime in which creation of particle is manifest. The same conclusion  can be made  in Ref \cite{Chair:2000vb} in which, pair production by a constant external field on NC space is also considered.   Note that $\Re_e\omega_{3+1}(x)=-\frac{m^4\wp}{16\pi}$ if $E=0$. For $B=0$, we use the taylor expansion \eqref{houklegrand} given in appendix \eqref{Appendix2}, and get
\bea
\Re_e\omega^{(B=0)}_{3+1}(x)&=&\frac{E^2}{4\pi^3}(1-\wp)\sum_{k=1}^\infty\frac{1}{k^2}\exp\Big[-\frac{k\pi m^2}{E}\Big]\cr
&-&\frac{m^4\wp}{16\pi}.
\eea 
The commutative limit which correspond to the case where $\wp=0$ is restored.

{\bf(3)}\,\, In the case of $1+1$  dimension we consider the electric field  ${\bf E}=E {\bf e}_x$, with $E>0$ and ${\bf B}={\bf 0}$. The nonvanishing component of the tensor $F_{\mu\nu}$ is given by $F_{01}=E$. The quantity $B(\theta, A, 1,1)$ given in Eq: \eqref{b} takes the form
\bea\label{somm}
&&B(\theta, A,1,1)=-\frac{m\theta E}{2}+\frac{i\theta\g^\mu}{2}\Big[iEA_\mu-E\partial_\mu\cr
&&-iA_\mu EX\partial_1-\partial_1 A_\mu\partial_0+\partial_\mu(EX)\partial_1\Big].
\eea
We remark that the result \eqref{somm} is also obtained by taking in \eqref{ko}  the magnetic field $B$  to be zero and by deleting the coordinates components $Y$ and $Z$.
Now, refer to \eqref{29} the Hermitian operator $\mathcal{X}(\theta)$ is obviously
\bea
\mathcal{X}_{H}(\theta)&=&\frac{\theta}{2}\Big[iE^2\g^0\g^1+2E^2\g^0\g^1 XP_1-4E^3 X^2\cr
&+&4E^2XP_0\Big].
\eea
We use the following results which are applicable in $ 1+ 1$  dimension:
\bea
\langle {\bf x}|e^{is(P+A)^2}|{\bf x}\rangle=\frac{E}{4\pi\sinh(Es)}
\eea
and
\bea
\langle {\bf x}|e^{isP^2}|{\bf x}\rangle=\frac{1}{4\pi s}.
\eea
By performing the same computation in appendix \eqref{Appendix1} and \eqref{Appendix2}, we come to
\bea\label{lanba}
\Re_{e}\omega^{(B=0)}_{1+1}(x)&=&\frac{E}{2\pi}(1-\wp)\sum_{k=1}^\infty\frac{1}{k}\exp\Big(-\frac{k\pi m^2}{E}\Big)\cr
&-&\frac{m^2\wp}{8}.
\eea
In the limit where $\theta\rightarrow 0 $ the equation \eqref{lanba}  restore the commutative limit given in \cite{Lin:1998rn}.

{\bf(4)}\,\,Let us discuss the case of $(2+1)$-dimensional space-time. The matrix $(\theta)^{\mu\nu}$  becomes singular and takes the form
\bea
\theta^{\mu\nu}=\left(\begin{array}{ccc}
0&\theta& 0\\
-\theta&0&0\\
0&0&0
\end{array}\right).
\eea
Then the noncommutativity of space-time is described by the commutation relation
\bea
[x^0, x^1]_\star=i\theta,\, [x^0, x^2]_\star=0,\,[x^1, x^2]_\star=0.
\eea
Despite this singularity of the matrix $(\theta^{\mu\nu})$, the SW application \eqref{expand}, \eqref{expand1} and \eqref{expand2}  are well satified. Therefore using  the fact that
\bea
\langle {\bf x}|e^{is(P+A)^2}|{\bf x}\rangle=\frac{(1-i)E}{2(2\pi)^{\frac{3}{2}}s^{\frac{1}{2}}\sinh(Es)}
\eea
and
\bea
\langle {\bf x}|e^{isP^2}|{\bf x}\rangle=\frac{1-i}{4(2\pi)^{\frac{3}{2}} s^{\frac{3}{2}}},
\eea
we can derive the probability density of particle creation from the vacuum by external constant EM fields. We obtain
\bea
\Re_{e}\omega^{(B=0)}_{2+1}(x)&=&\frac{E^{\frac{3}{2}}}{4\pi^2}(1-\wp)\sum_{k=1}^\infty\frac{1}{k^{\frac{3}{2}}}\exp\Big(-\frac{k\pi m^2}{E}\Big)\cr
&-&\frac{m^3\wp}{8(2\pi)^{\frac{1}{2}}}
\eea
and obviously the limit $\theta=0$ restore the result of Ref: \cite{Lin:1998rn}.

{\bf(5)}\,\,Furthermore, the previous investigation \cite{Hounkonnou:2000im},  devoted to such EM field as ${\bf E}=E\cos (t){\bf e_x}$ and ${\bf B}=B{\bf e_x},$ has been also considered here in the framework of  the
NCFT. Indeed, following step by step  the approach displayed earlier in this work,  and using the following relation:
\bea
\langle {\bf x}|e^{is(P+A)^2}|{\bf x}\rangle=\frac{-iEB\cos (t) }{16\pi^2\sinh(Es)\sin(Bs)},
\eea 
after some algebra, we get
\bea
&&\widetilde\omega_{3+1}(t)=\frac{1}{4\pi^2i}\int_{0}^\infty\, ds\,\frac{e^{ism^2}}{s}\Bigg[\Big(1-\wp\Big)\cr
&&\times \frac{EB\cos(t)\cosh[E\cos(t)s]\cot(Bs)}{\sinh(Es)}-\frac{1}{s^2}\Bigg].
\eea
We found that the  probability density of the pair production of Dirac particle in NC space-time with alternating EM field is given by 
\bea\label{intfant120}
&&\Re_e\widetilde \omega_{3+1}(t)=-\frac{m^4\wp}{16\pi}+\frac{EB}{4\pi^2}\big(1-\wp\big)\cr
&\times&\sum_{n=1}^\infty\frac{(-1)^n}{n}\cos\big(n\pi\cos(t)\big)\coth\Big(n\pi\frac{B}{E}\Big)\cr
&\times&\exp\Big(-\frac{n\pi m^2}{E}\Big),
\eea
from which, 
 in the limit where the NC parameter $\theta=0$,  we recover the  formula
of Hounkonnou et al 
\cite{Hounkonnou:2000im} i.e.
\bea\label{intfant122}
\widetilde \omega(t)&=&
\frac{EB}{4\pi^2}\sum_{n=1}^\infty\frac{(-1)^n}{n}\cos\big(n\pi\cos(t)\big)\coth\Big(n\pi\frac{B}{E}\Big)\cr
&&\times\exp\Big(-\frac{n\pi m^2}{E}\Big).
\eea
A more compact form of the relation \eqref{intfant120} in the case of arbitrary
$D=d+1$-dimensions can be also given in the same way. We get for $B=0$ the following results
\bea\label{houknas}
\widetilde\omega_{d+1}(t)&=&\big(1-\wp\big)\frac{E^{\frac{d+1}{2}}\cos(t)}{(2\pi)^d}\sum_{n=1}^\infty\,\frac{(-1)^n}{n^{\frac{d+1}{2}}}\cr
&\times&\cos\Big(\frac{n\pi}{\cos(t)}\Big)\exp\Big(-\frac{n\pi m^2}{E}\Big)\cr
&-&\frac{m^{d+1}\wp}{4(2)^{\frac{d+1}{2}}(\pi)^{\frac{d-1}{2}}}.
\eea
Also, by replacing the vector field $A_\mu$ by $\mathcal A_\mu=A_\mu+f_\mu$, where $A_\mu=(-EX,0,0,0)$ and $f_\mu=(-E\sin(x),0,0,0)$ corresponding to the plane wave function, we get
\bea\label{houknas1}
\widetilde\omega_{d+1}(x)&=&\big(1-\wp\big)\frac{(2E)^{\frac{d+1}{2}}\big(1+\cos(x))}{(2\pi)^d}\sum_{n=1}^\infty\,\frac{(-1)^n}{n^{\frac{d+1}{2}}}\cr
&\times&\cos\Big(n\pi\frac{1+cos(x)}{2}\Big)\exp\Big(-\frac{n\pi m^2}{2E}\Big)\cr
&-&\frac{m^{d+1}\wp}{4(2)^{\frac{d+1}{2}}(\pi)^{\frac{d-1}{2}}}.
\eea
All these results use the computations performed in
the Appendices \eqref{Appendix1} and \eqref{Appendix2}. 
In the limit where $\theta=0$, the relations  \eqref{houknas} and \eqref{houknas1} lead  to   the results of  \cite{Hounkonnou:2000im}. See also \cite{Gavrilov:1996pz} in which the  generalization in arbitrary dimensions is well given in the case of quasiconstant external EM fields.

\section{Concluding remarks}\label{sec4}
In this paper, we have considered NC theory of fermionic field  interacting with its corresponding boson. We have used the Seiberg-Witten expansion  describing the relation between the NC and commutative variables, to compute the probability density of pair production of  NC fermions. 
%The technical computations that we have  provided,  allowed us to
 We have showed that,  in the limit where the NC parameter $\theta=0,$ we recover the result of Qiong-Gui Lin \cite{Lin:1998rn}. Our study has highlighted that  the noncommutativity of space-time increases the density $\omega$ of the probability of pair creation of the fermion particle. Our results can be easily extended to take into account
%The discussion of our result in the particular cases where $B=0$, and where $B=E$ can also be made.
%Finally 
the case where  $D=1+1.$ 
%take form after simple computation.

 \section*{Acknowledgements} 
The authors thank Matteo Smerlak for fruitful discussions and comments that allowed to improve the paper. We also  thank all the referees, who have anonymously contributed to the improvement of this document.

D. O. S. research is supported in part by the Perimeter Institute for Theoretical Physics (Waterloo) and by the Fields Institute for Research in Mathematical Sciences (Toronto).
Research at the Perimeter Institute is supported by the Government of Canada through Industry Canada and by the Province of Ontario through the Ministry of Economic Development \& Innovation.

E. E. N.  and M. N. H  works are partially supported by the Abdus Salam International Centre for Theoretical Physics (ICTP, Trieste, Italy) through the Office of External Activities (OEA)-Prj-15. The ICMPA is in partnership with the Daniel Iagolnitzer Foundation (DIF), France.

%%%%%%%%%%%%%%%%%%%%%%%%%%%%%%
\appendix

%\begin{center}
%{\bf APPENDIX}\\
%
%\end{center}

%In Appendix 1,
\section{Proof of  Theorem 1}
\label{Appendix1}
We give the proof of  Theorem \eqref{th1}.
%\eqref{th1}.
We have introduced   the two quantities: $\theta_0$ and $\wp$,  and decompose the NC parameter $\theta$ as 
\beq
\theta=\wp\,\theta_0,\,\,\mbox{ such that}\,\,\,\theta_0<<1.
\eeq
The first step is to re-express $\mathcal O_{nc}(\theta)$ as follows:
\begin{widetext}
\bea
\mathcal O_{nc}(\theta)&=&e^{\frac{is}{2}\sigma^{\mu\nu}F_{\mu\nu}}\int\, d{\bf y}\,\langle {\bf x} |e^{is(P+A)^2}|{\bf y}\rangle
\langle {\bf y}|-is\big[m^2\theta(B-E)-\mathcal{X}_H(\theta)\big]|{\bf x} \rangle.\cr
&=&\wp\,e^{\frac{is}{2}\sigma^{\mu\nu}F_{\mu\nu}}\int\, d{\bf k}\,d{\bf y}\,\langle {\bf x} |e^{is(P+A)^2}|{\bf y}\rangle\langle {\bf y}|is\theta_0\mathcal G(E,B,\theta)|{\bf k}\rangle\langle {\bf k}|{\bf x}\rangle\cr
&=&\wp\,e^{\frac{is}{2}\sigma^{\mu\nu}F_{\mu\nu}}\int\, d{\bf k}\,d{\bf x}\,\langle {\bf x} |e^{is(P+A)^2}|{\bf x}\rangle \big( is\theta_0\mathcal G(E,B,\theta)\big),
\eea
where
\bea\label{above}
\mathcal G(E,B,\theta)&=&\Bigg\{m^2(E-B)+\frac{1}{2}\Big[iEB\g^{3}\g^{2}
+iE^{2}\g^{0}\g^{1}+iB^{2}\g^{3}\g^{2}+\frac{1}{2} i(\g^{0}\g^{1}+\g^3\g^2)EB+ \g^{0}\g^{1}EByk_{2}\cr 
& +&
2E^2\g^0\g^1xk_1+\g^{0}\g^{3}(E^{2}B-EB^2)xy-\g^{1}\g^{3}EByk_{1}-(4E^{3}+3BE^2)x^{2}
+(2B^{3}+B^2E)y^{2}
\cr
&+&(4E^2+5EB)xk_{0}
+(2B^2+3EB)yk_3-2Bk_{0}^{2}
+2Bk_1^2
+2Ek^{2}_{2}+2Ek^{2}_{3}
\Big]\Bigg\}.
\eea
\end{widetext}
The  expression \eqref{above} is subdivided into three contributions, namely
\bea
\mathcal G_0&=&m^2(E-B)+\frac{1}{2}\Big[iEB\g^{3}\g^{2}
+iE^{2}\g^{0}\g^{1}\cr
&+&iB^{2}\g^{3}\g^{2}+\frac{1}{2} i(\g^{0}\g^{1}+\g^3\g^2)EB\Big],
\eea
\bea
\mathcal G_1&=&\frac{1}{2}\Big[\g^{0}\g^{1}EByk_{2} +
2E^2\g^0\g^1xk_1-\g^{1}\g^{3}EByk_{1}\cr
&+&(4E^2+5EB)xk_{0}
+(2B^2+3EB)yk_3-2Bk_{0}^{2}
\cr
&&
+2Bk_1^2
+2Ek^{2}_{2}+2Ek^{2}_{3}\Big]
\eea
and
\bea
\mathcal G_2&=&\frac{1}{2}\Big[\g^{0}\g^{3}(E^{2}B-EB^2)xy-(4E^{3}+3BE^2)x^{2}\cr
&+&(2B^{3}+B^2E)y^{2}\Big].
\eea
We are especially interested in the first order Taylor expansion of the form:
\bea
is\theta_0\mathcal G(E,B,\theta)\equiv \exp(is\theta_0\mathcal G(E,B,\theta))-1.
\eea
Coming back to equation \eqref{3400}, this equation can be expressed as
\bea\label{Useful}
&&\mathcal O=(1-\wp)\mathcal O_c+\wp\,e^{\frac{is}{2}\sigma^{\mu\nu}F_{\mu\nu}}\cr
&&\times\int\, d{\bf k}\,d{\bf x}\,\langle {\bf x} |e^{is(P+A)^2}|{\bf x}\rangle \exp\Big( is\theta_0\mathcal G(E,B,\theta)\Big).\cr&&
\eea
 First we consider  the contribution  $\exp\big( is\theta_0\mathcal G_1\big)$ of $\exp\Big( is\theta_0\mathcal G(E,B,\theta)\Big)$ in \eqref{Useful} and  the integral relation
\bea
\int_{-\infty}^{\infty}\,e^{isx^2}dx=\begin{cases}e^{\frac{\pi}{4}}\sqrt{\frac{\pi}{s}}& \text{ for } s>0\\
e^{-\frac{\pi}{4}}\sqrt{\frac{\pi}{s}}& \text{ for } s<0
\end{cases}.
\eea
We  get, respectively,
\bea\label{k1}
&&\mathcal K_1=\int\,dk_0\exp\Big[ \frac{is\theta_0}{2}\Big(-2Bk_0^2\cr
&&+(4E^2+5EB)xk_0\Big)\Big]\cr
&&=e^{-\frac{\pi}{4}}\sqrt{\frac{\pi}{sB\theta_0}}\exp\Big[\frac{is\theta_0}{16B}(4E^2+5EB)^2x^2\Big]\cr&&
\eea
\bea\label{k2}
\mathcal K_2&=&\int dk_1\exp\Big[ \frac{is\theta_0}{2}\Big(2Bk_1^2+2\g^0\g^1E^2xk_1\cr
&-&\g^1\g^3EByk_1\Big)
\Big]\cr
&=&e^{\frac{\pi}{4}}\sqrt{\frac{\pi}{sB\theta_0}}\exp\Big[-\frac{is\theta_0}{16B}(2\g^0\g^1E^2x\cr
&-&\g^1\g^3EBy)^2\Big]
\eea
\bea\label{k3}
\mathcal K_3&=&\int\,dk_1\exp\Big[ \frac{is\theta_0}{2}\Big(2Ek_2^2+\g^0\g^1EByk_2\Big)
\Big]\cr
&=&e^{\frac{\pi}{4}}\sqrt{\frac{\pi}{sE\theta_0}}\exp\Big[\frac{is\theta_0}{16E}E^2B^2y^2\Big]
\eea
\bea\label{k4}
\mathcal K_4&=&\int\,dk_3\exp\Big[ \frac{is\theta_0}{2}\Big(2Ek_3^2\cr
&+&(2B^2+3EB)yk_3\Big)\Big]\cr
&=&e^{\frac{\pi}{4}}\sqrt{\frac{\pi}{sE\theta_0}}\exp\Big[\frac{is\theta_0}{16E}(2B^2+3EB)^2y^2\Big].\cr&&
\eea

Using the properties of the gamma matrices and the results of \cite{Lin:1998rn} and  \cite{Hounkonnou:2000im} we get the following: 
\bea
\tr e^{\frac{is}{2}\sigma^{\mu\nu}F_{\mu\nu}}=4\cosh(Es)\cos(Bs)
\eea
\bea
\langle {\bf x} |e^{is(P+A)^2}|{\bf x}\rangle=-\frac{iEB}{16\pi^2\sinh(Es)\sin(Bs)}
\eea
\beq
\langle x|e^{isP^2}|x \rangle=-\frac{i}{16\pi^2 s^2}.
\eeq
We can evaluate the trace of  relevant quantities  in equation \eqref{Useful}. On the other hand,
before  obtaining \eqref{C}, we compute  the trace of 
$
\int \,d{\bf k}\,\exp\big[ is\theta_0\langle {\bf k}|\mathcal G(E,B,\theta)|{\bf k}\rangle \big],
$ i.e.
\bea\label{dense}
&&\int\,d{\bf x} \int \,d{\bf k}\,\exp\big[ is\theta_0
\langle {\bf k}|\mathcal G(E,B,\theta)| {\bf k}\rangle\big]\cr
&&=\exp\Big[is\theta_0\mathcal G_0\Big]\int\,dtdx\,dydz \,\prod_{j=1}^4
\mathcal K_j \exp\Big[is\theta_0\mathcal G_2\Big].\cr&&
\eea
This is  obtained by using  the Gaussian integral. Moreover, taking  into account the equations \eqref{k1}, \eqref{k2}, \eqref{k3} and \eqref{k4} we get:
\begin{widetext}
\bea
\prod_{j=1}^4\mathcal K_j \exp\Big[is\theta_0\mathcal G_2\Big]&=&\frac{i\pi^2}{s^2\theta_0^{2}EB}\exp\Bigg\{\frac{is\theta_0}{16B}(4E^2+5EB)^2x^2-\frac{is\theta_0}{16B}(2\g^0\g^1E^2x-\g^1\g^3EBy)^2\cr
&+&\frac{is\theta_0}{16E}E^2B^2y^2+\frac{is\theta_0}{16E}(2B^2+3EB)^2y^2+\frac{is\theta_0}{2}\Big[\g^{0}\g^{3}(E^{2}B-EB^2)xy\cr
&-&(4E^{3}+3BE^2)x^{2}+(2B^{3}+B^2E)y^{2}\Big]\bigg\}.
\eea
In this relation, the quantity 
\bea
&&\exp\Bigg\{\frac{is\theta_0}{16B}\Big[(4E^2+5EB)^2x^2-4E^4x^2+8B\g^{0}\g^{3}(E^{2}B-EB^2)xy-8B(4E^{3}+3BE^2)x^{2}\Big]\Bigg\}
\eea
 contributes to the integration respect to $x$. We get after integration:
\bea\label{trr}
e^{\frac{i\pi}{4}}\sqrt{\frac{16B\pi}{s\theta_0(12 E^4+8E^3B+E^2B^2)}}\exp\Big[is\theta_0\frac{B(E^{2}B-EB^2)^2y^2}{(12 E^4+8E^3B+E^2B^2)}\Big].
\eea
Consider \eqref{trr}, the expression  
\bea
&&\exp\Bigg\{is\theta_0\frac{B(E^{2}B-EB^2)^2y^2}{(12 E^4+8E^3B+E^2B^2)}+\frac{is\theta_0}{16B}E^2B^2y^2+\frac{is\theta_0}{16E}E^2B^2y^2+\frac{is\theta_0}{16E}(2B^2+3EB)^2y^2\cr
&&+\frac{is\theta_0}{2}(2B^{3}+B^2E)y^{2}\Bigg\}\cr
&&=\exp\Big\{is\theta_0\frac{E(4B^6+76EB^5+258 E^2 B^4+494 E^3
B^3 +224 E^4 B^2+12E^5B)y^2}{16(12 E^4+8E^3B+E^2B^2)}\Big\}
\eea
 contributes to the integration with respect to $y$. We get
after integration:
\bea
e^{\frac{i\pi}{4}}\sqrt{\frac{16\pi(12 E^4+8E^3B+E^2B^2)}{s\theta_0E(4B^6+76EB^5+258 E^2 B^4+494 E^3
B^3 +224 E^4 B^2+12E^5B)}}.
\eea
Consider $f(E,B)$ as positive defined function given by the following:
\bea
f(E,B)=\Big[4B^6+76EB^5+258 E^2 B^4+494 E^3
B^3 +224 E^4 B^2+12E^5B\Big]^{-1}.
\eea 
Hence, it is straightforward to obtain
\bea
\int\,dx\,dy \,\prod_{j=1}^4
\mathcal K_j \exp\Big[is\theta_0\mathcal G_2\Big]=-\frac{16\pi^3}{s^3EB\theta_0^{3}}\sqrt{\frac{B}{E}f(E,B)}.
\eea
\end{widetext}
  We choose the subset $M\in\mathbb{R}^2$ such that $\int \frac{dt}{t_0} dz=b$, ( For example $M$  could be the unit disk of $\mathbb{R}^2$). We come to
\bea
&&\int\,dt\,dz\,dx\,dy \int \,d{\bf k}\,\big[ is\theta_0
\langle {\bf k}|\mathcal G(E,B,\theta)| {\bf k}\rangle\big]\cr
&&=-\frac{16\pi^3 b}{s^3EB\theta_0^{3}}\sqrt{\frac{B}{E}f(E,B)}\exp\Big[is\theta_0\mathcal G_0\Big],\cr&&
\eea
with $\int \frac{dt}{t_0} dz=b\equiv[L^2]$. Finally the
following expression is obviously satisfied:
\bea\label{Useful2}
\tr\mathcal O=
\Big(1-\wp-\sigma(\theta_0,E,B)\Big)\tr\mathcal O_c,
\eea
where
\bea
\tr \mathcal O_c=\frac{1}{4\pi^2 i}EB\cosh(Es)\cot(Bs).
\eea
This ends the proof of Theorem 1.
%\eqref{th1}. 
\section{Proof of  Theorem 2} \label{Appendix2}
This section is devoted to the proof of  Theorem \eqref{th2}.
To evaluate the integral \eqref{intfan} before getting \eqref{intfant1}, we need to collect  information about physical property in the  limit where the magnetic field $B$ tends to zero. This is clearly given in \cite{Lin:1998rn}. However, we
think that it may be instructive to collect here all the arguments and rewrite the complete
proof for our purpose.  
All the integral will be performed in the half 
 complex plane. We will select only the positive half plane.  Consider first the integral $\int_{-\infty}^{\infty}\,ds\,\frac{e^{ism^2}}{s^3}$.  Using the residue theorem, we  simply get
\bea
\int_{-\infty}^{\infty}\,ds\,\frac{e^{ism^2}}{s^3}=-i\pi\frac{m^4}{2}.
\eea
Let us consider $\int_{-\infty}^{\infty}\,ds\,\frac{e^{ism^2}}{s}\coth(Es)\cot(Bs)$. Let $h(z)=\frac{e^{izm^2}}{z}\coth(Ez)\cot(Bz),\,z\in\mathbb{C}.$ The integrand has singularities at point $z=0$ (poles of order $3$), at $z=\frac{in\pi}{E}$ and $z=\frac{n\pi}{B}$ (simple poles).  Let $Res(z_0)$ be the residue of $h(z)$ at the point $z_0\in\mathbb{C}$. 
We  write the Taylor expansion of  $\cot(z)$ and $\coth(z)$ at point $z_0$ as
\bea\label{dinoeno}
\cot(z)&=&\frac{1}{z-z_0}+\sum_{k=1}^\infty\,(-1)^k 2^{2k}\frac{B_{2k}}{(2k)!}(z-z_0)^{2k-1}\cr
&=&\frac{1}{z-z_0}-\frac{z-z_0}{3}-\frac{(z-z_0)^3}{45}+\cdots
\eea
and
\bea\label{houklegrand}
\coth(z)&=&\frac{1}{z-z_0}+\sum_{k=1}^\infty\, 2^{2k}\frac{B_{2k}}{(2k)!}(z-z_0)^{2k-1}\cr
&=&\frac{1}{z-z_0}+\frac{z-z_0}{3}-\frac{(z-z_0)^3}{45}+\cdots
\eea
where $B_n$  stands for the Bernoulli numbers with the initial values $(B_0=1,\,B_1=-1/2,\,B_2=1/6,\, B_4=-1/30,\, \, B_{2n-1}=0,n=2,3,\cdots)$. After  taking into account  the Taylor expansion of $\coth(Ez)$ $\cot(Bz)$ in the equations \eqref{dinoeno} and \eqref{houklegrand}, and the fact that $\cot (iz)=-i\tanh(z)$,  we get simply
\beq\label{danger0}
Res(0)=-\frac{m^4}{2BE},
\eeq

\beq\label{danger}
Res\Big(\frac{n\pi}{B}\Big)=\frac{1}{n\pi}\exp\Big(im^2\frac{n\pi}{B}\Big)\coth\Big(\frac{n\pi E}{B}\Big)
\eeq
and
\beq\label{danger1}
Res\Big(\frac{in\pi}{E}\Big)=-\frac{1}{\pi n}\exp\Big(-\frac{n\pi m^2}{E}\Big)\coth\Big(\frac{n\pi B}{E}\Big).
\eeq
Then
\bea\label{math}
&&\int_{-\infty}^{\infty}\,ds\,\frac{e^{ism^2}}{s}\coth(Es)\cot(Bs)\cr
&&=2i\pi\sum_{n=1}^\infty \Big[\frac{1}{n\pi}\exp\Big(im^2\frac{n\pi}{B}\Big)\coth\Big(\frac{n\pi E}{B}\Big)\cr
&&-\frac{1}{n\pi}\exp\Big(-\frac{n\pi m^2}{E}\Big)\coth\Big(\frac{n\pi B}{E}\Big)\Big]\cr
&&-i\pi\frac{m^4}{2BE}
\eea
By multiplying the above result by $EB$ it is clear that the limit $B\rightarrow 0$ is not well defined. This is why $\eqref{danger}$
cannot be taken into account in the physical situation. Therefore
 \eqref{math}  reduces to
 \bea\label{phys}
&&\int_{-\infty}^{\infty}\,ds\,\frac{e^{ism^2}}{s}\coth(Es)\cot(Bs)\cr
&&=-i\pi\frac{m^4}{2BE}-2i\pi\sum_{n=1}^\infty \Big[\frac{1}{n\pi}e^{-\frac{n\pi m^2}{E}}\coth\Big(\frac{n\pi B}{E}\Big)\Big].\cr
&&
\eea
Now let $k(z)=\frac{e^{izm^2}}{z^4}\coth(Ez)\cot(Bz),\,z\in\mathbb{C}.$ The integrand has singularities at point $z=0$ (poles of order $6$), at $z=\frac{in\pi}{E}$ and $z=\frac{n\pi}{B}$ (simple poles). Using the same argument like \eqref{danger0}, \eqref{danger} and \eqref{danger1} we get, respectively,
\beq\label{danger00}
Res(0)=\frac{im^{10}}{120BE},
\eeq

\beq\label{dangerr}
Res\Big(\frac{n\pi}{B}\Big)=\frac{B^3}{(n\pi)^4}\exp\Big(im^2\frac{n\pi}{B}\Big)\coth\Big(\frac{n\pi E}{B}\Big)
\eeq
and
\beq\label{danger11}
Res\Big(\frac{in\pi}{E}\Big)=\frac{iE^3}{(\pi n)^4}\exp\Big(-\frac{n\pi m^2}{E}\Big)\coth\Big(\frac{n\pi B}{E}\Big).
\eeq
Now we come to the interpretation of the equations \eqref{danger00}, \eqref{dangerr} and \eqref{danger11}.
\begin{itemize}
\item The equations \eqref{danger00} and \eqref{danger11} lead to a complex probability density and then cannot be taken into account.
\item As we have seen in \eqref{danger}, the equation  \eqref{dangerr} leads to  a singularity at the limit 
$B\rightarrow 0$. This pointless expression  also will not  contribute to $\Re_e(\omega).$
\end{itemize}
The same analysis can be provided, using the holomorphic function $$\frac{\coth(Ez)\cot(Bz)}{z^4}\exp\Big[iz(m^2+\theta_0\mathcal G_0)\Big].$$
Taking into account the above two comments, we conclude that the function 
$
\sigma(\theta_0,E,B)
$
 do not contribute to the probality density, and therefore the parameter $\theta_0$ disappears in the equation \eqref{intfant1}. 
Finally, by taking into account only the relation \eqref{phys}, the Theorem 
\eqref{th2}
 is  proved. 
%\beq
%\int_0^\infty\,ds\, s^me^{is^n}=\frac{1}{n!}\Gamma\Big(\frac{m+1}{n}\Big) e^{\frac{i\pi(m+1)}{2n}}
%\eeq

%%%%%%%%%%%%%%%%%%%%%%
%%%%%%%%%%%%%%%%%%%%%%%%%%%%%%%%%%%%%%%%%%%%%%%%%%%%


\begin{thebibliography}{99}
% \expandafter\ifx\csname natexlab\endcsname\relax\def\natexlab#1{#1}\fi
%\expandafter\ifx\csname bibnamefont\endcsname\relax
%  \def\bibnamefont#1{#1}\fi
%\expandafter\ifx\csname bibfnamefont\endcsname\relax
%  \def\bibfnamefont#1{#1}\fi
%\expandafter\ifx\csname citenamefont\endcsname\relax
%  \def\citenamefont#1{#1}\fi
%\expandafter\ifx\csname url\endcsname\relax
%  \def\url#1{\texttt{#1}}\fi
%\expandafter\ifx\csname urlprefix\endcsname\relax\def\urlprefix{URL }\fi
%\providecommand{\bibinfo}[2]{#2}
%\providecommand{\eprint}[2][]{\url{#2}}

\bibitem{Snyder:1946qz} 
  H.~S.~Snyder,
  ``Quantized space-time,''
  Phys.\ Rev.\  {\bf 71}, 38 (1947).
  %%CITATION = PHRVA,71,38;%%
  %1071 citations counted in INSPIRE as of 22 Mar 2014

%\cite{Connes:1994yd}
\bibitem{Connes:1994yd} 
  A.~Connes,
  ``Noncommutative Geometry,''
  ISBN-9780121858605.
  %%CITATION = ISBN-9780121858605;%%
  %24 citations counted in INSPIRE as of 23 May 2014

%\cite{Connes:1990qp}
\bibitem{Connes:1990qp} 
  A.~Connes and J.~Lott,
  %``Particle Models and Noncommutative Geometry (Expanded Version),''
  Nucl.\ Phys.\ Proc.\ Suppl.\  {\bf 18B}, 29 (1991).
  %%CITATION = NUPHZ,18B,29;%%
  %229 citations counted in INSPIRE as of 23 May 2014

%\cite{Woronowicz:1987wr}
\bibitem{Woronowicz:1987wr} 
  S.~L.~Woronowicz,
  %``Twisted SU(2) group: An Example of a noncommutative differential calculus,''
  Publ.\ Res.\ Inst.\ Math.\ Sci.\ Kyoto {\bf 23}, 117 (1987).
  %%CITATION = PBMAA,23,117;%%
  %225 citations counted in INSPIRE as of 24 May 2014

%\cite{Doplicher:1994tu}
\bibitem{Doplicher:1994tu} 
  S.~Doplicher, K.~Fredenhagen and J.~E.~Roberts,
  %``The Quantum structure of space-time at the Planck scale and quantum fields,''
  Commun.\ Math.\ Phys.\  {\bf 172}, 187 (1995)
  [hep-th/0303037].
  %%CITATION = HEP-TH/0303037;%%
  %725 citations counted in INSPIRE as of 19 May 2014

%\cite{Majid:1999tc}
\bibitem{Majid:1999tc} 
  S.~Majid,
%  ``Meaning of noncommutative geometry and the Planck scale quantum group,''
  Lect.\ Notes Phys.\  {\bf 541}, 227 (2000)
  [hep-th/0006166].
  %%CITATION = HEP-TH/0006166;%%
  %31 citations counted in INSPIRE as of 19 May 2014

%\cite{Szabo:2001kg}
\bibitem{Szabo:2001kg} 
  R.~J.~Szabo,
  %``Quantum field theory on noncommutative spaces,''
  Phys.\ Rept.\  {\bf 378}, 207 (2003)
  [hep-th/0109162].
  %%CITATION = HEP-TH/0109162;%%
  %1039 citations counted in INSPIRE as of 24 May 2014

%\cite{Banburski:2013jfa}
\bibitem{Banburski:2013jfa} 
  A.~Banburski and L.~Freidel,
 % ``Snyder Momentum Space in Relative Locality,''
  arXiv:1308.0300 [gr-qc].
  %%CITATION = ARXIV:1308.0300;%%

%\cite{Ferrari:2005ng}
\bibitem{Ferrari:2005ng} 
  A.~F.~Ferrari, H.~O.~Girotti and M.~Gomes,
%  ``Lorentz symmetry breaking in the noncommutative Wess-Zumino model: One loop corrections,''
  Phys.\ Rev.\ D {\bf 73}, 047703 (2006)
  [hep-th/0510108].
  %%CITATION = HEP-TH/0510108;%%
  %3 citations counted in INSPIRE as of 19 May 2014  %3 citations counted in INSPIRE as of 19 May 2014

%\cite{Imai:2000kq}
\bibitem{Imai:2000kq} 
  S.~'I.~Imai and N.~Sasakura,
 % ``Scalar field theories in a Lorentz invariant three-dimensional noncommutative space-time,''
  JHEP {\bf 0009}, 032 (2000)
  [hep-th/0005178].
  %%CITATION = HEP-TH/0005178;%%
  %22 citations counted in INSPIRE as of 19 May 2014

%\cite{Carmona:2002iv}
\bibitem{Carmona:2002iv} 
  J.~M.~Carmona, J.~L.~Cortes, J.~Gamboa and F.~Mendez,
 % ``Noncommutativity in field space and Lorentz invariance violation,''
  Phys.\ Lett.\ B {\bf 565}, 222 (2003)
  [hep-th/0207158].
  %%CITATION = HEP-TH/0207158;%%
  %78 citations counted in INSPIRE as of 19 May 2014

%\cite{Jackiw:2001dj}
\bibitem{Jackiw:2001dj} 
  R.~Jackiw,
  %``Physical instances of noncommuting coordinates,''
  Nucl.\ Phys.\ Proc.\ Suppl.\  {\bf 108}, 30 (2002)
  [Phys.\ Part.\ Nucl.\  {\bf 33}, S6 (2002)]
  [Lect.\ Notes Phys.\  {\bf 616}, 294 (2003)]
  [hep-th/0110057].
  %%CITATION = HEP-TH/0110057;%%
  %104 citations counted in INSPIRE as of 26 May 2014
  %
  %
  %
  %%\cite{Berrino:2002ss}
\bibitem{Berrino:2002ss} 
  G.~Berrino, S.~L.~Cacciatori, A.~Celi, L.~Martucci and A.~Vicini,
  %``Noncommutative electrodynamics,''
  Phys.\ Rev.\ D {\bf 67}, 065021 (2003)
  [hep-th/0210171].
  %%CITATION = HEP-TH/0210171;%%
  %22 citations counted in INSPIRE as of 26 May 2014

%\cite{Raasakka:2010ev}
\bibitem{Raasakka:2010ev} 
  M.~Raasakka and A.~Tureanu,
 % ``On UV/IR Mixing via Seiberg-Witten Map for Noncommutative QED,''
  Phys.\ Rev.\ D {\bf 81}, 125004 (2010)
  [arXiv:1002.4531 [hep-th]].
  %%CITATION = ARXIV:1002.4531;%%
  %5 citations counted in INSPIRE as of 19 May 2014

%\cite{Liao:2002kd}
\bibitem{Liao:2002kd} 
  Y.~Liao and C.~Dehne,
%  ``Some phenomenological consequences of the time ordered perturbation theory of QED on noncommutative space-time,''
  Eur.\ Phys.\ J.\ C {\bf 29}, 125 (2003)
  [hep-ph/0211425].
  %%CITATION = HEP-PH/0211425;%%
  %23 citations counted in INSPIRE as of 19 May 2014

%\cite{Seiberg:1999vs}
\bibitem{Seiberg:1999vs} 
  N.~Seiberg and E.~Witten,
  %``String theory and noncommutative geometry,''
  JHEP {\bf 9909}, 032 (1999)
  [hep-th/9908142].
  %%CITATION = HEP-TH/9908142;%%
  %3253 citations counted in INSPIRE as of 22 Mar 2014

%\cite{Madore:2000en}
\bibitem{Madore:2000en} 
  J.~Madore, S.~Schraml, P.~Schupp and J.~Wess,
  %``Gauge theory on noncommutative spaces,''
  Eur.\ Phys.\ J.\ C {\bf 16}, 161 (2000)
  [hep-th/0001203].
  %%CITATION = HEP-TH/0001203;%%
  %443 citations counted in INSPIRE as of 22 May 2014

%\cite{Madore:1997dt}
\bibitem{Madore:1997dt} 
  J.~Madore and L.~A.~Saeger,
  %``Topology at the Planck length,''
  Class.\ Quant.\ Grav.\  {\bf 15}, 811 (1998)
  [gr-qc/9708053].
  %%CITATION = GR-QC/9708053;%%
  %23 citations counted in INSPIRE as of 22 May 2014


%\cite{Fidanza:2001qm}
\bibitem{Fidanza:2001qm} 
  S.~Fidanza,
  %``Towards an explicit expression of the Seiberg-Witten map at all orders,''
  JHEP {\bf 0206}, 016 (2002)
  [hep-th/0112027].
  %%CITATION = HEP-TH/0112027;%%
  %20 citations counted in INSPIRE as of 22 May 2014


%\cite{Ulker:2012yk}
\bibitem{Ulker:2012yk} 
  K.~Ulker,
  %``On the All Order Solutions of Seiberg-Witten Map for Noncommutative Gauge Theories,''
  Int.\ J.\ Mod.\ Phys.\ Conf.\ Ser.\  {\bf 13}, 191 (2012)
  [arXiv:1201.2192 [hep-th]].
  %%CITATION = ARXIV:1201.2192;%%

%\cite{Ulker:2007fm}
\bibitem{Ulker:2007fm} 
  K.~Ulker and B.~Yapiskan,
  %``Seiberg-Witten maps to all orders,''
  Phys.\ Rev.\ D {\bf 77}, 065006 (2008)
  [arXiv:0712.0506 [hep-th]].
  %%CITATION = ARXIV:0712.0506;%%
  %20 citations counted in INSPIRE as of 22 May 2014



%\cite{Bozkaya:2002at}
\bibitem{Bozkaya:2002at} 
  H.~Bozkaya, P.~Fischer, H.~Grosse, M.~Pitschmann, V.~Putz, M.~Schweda and R.~Wulkenhaar,
 % ``Space-time noncommutative field theories and causality,''
  Eur.\ Phys.\ J.\ C {\bf 29}, 133 (2003)
  [hep-th/0209253].
  %%CITATION = HEP-TH/0209253;%%
  %40 citations counted in INSPIRE as of 19 May 2014

%\cite{Harms:2006dv}
\bibitem{Harms:2006dv} 
  B.~Harms and O.~Micu,
  %``Noncommutative quantum Hall effect and Aharonov-Bohm effect,''
  J.\ Phys.\ A {\bf 40}, 10337 (2007)
  [hep-th/0610081].
  %%CITATION = HEP-TH/0610081;%%
  %15 citations counted in INSPIRE as of 25 May 2014

%\cite{Scholtz:2005vg}
\bibitem{Scholtz:2005vg} 
  F.~G.~Scholtz, B.~Chakraborty, S.~Gangopadhyay and J.~Govaerts,
  %``Interactions and non-commutativity in quantum Hall systems,''
  J.\ Phys.\ A {\bf 38}, 9849 (2005)
  [cond-mat/0509331 [cond-mat.mes-hall]].
  %%CITATION = COND-MAT/0509331;%%
  %33 citations counted in INSPIRE as of 25 May 2014



%\cite{Schwinger:1948iu}
\bibitem{Schwinger:1948iu} 
  J.~S.~Schwinger,
  %``On Quantum electrodynamics and the magnetic moment of the electron,''
  Phys.\ Rev.\  {\bf 73}, 416 (1948).
  %%CITATION = PHRVA,73,416;%%
  %388 citations counted in INSPIRE as of 13 May 2014
%%\cite{Schwinger:1951xk}
%\bibitem{Schwinger:1951xk} 
%  J.~S.~Schwinger,
%  ``The Theory of quantized fields. 1.,''
%  Phys.\ Rev.\  {\bf 82}, 914 (1951).
%  %%CITATION = PHRVA,82,914;%%
%  %381 citations counted in INSPIRE as of 13 May 2014
%
%%\cite{Schwinger:1953tb}
%\bibitem{Schwinger:1953tb} 
%  J.~S.~Schwinger,
%  ``The Theory of quantized fields. 2.,''
%  Phys.\ Rev.\  {\bf 91}, 713 (1953).
%  %%CITATION = PHRVA,91,713;%%
%  %209 citations counted in INSPIRE as of 13 May 2014
%
%
%
%%\cite{Schwinger:1953zza}
%\bibitem{Schwinger:1953zza} 
%  J.~Schwinger,
%  ``The Theory of Quantized Fields. III,''
%  Phys.\ Rev.\  {\bf 91}, 728 (1953).
%  %%CITATION = PHRVA,91,728;%%
%  %66 citations counted in INSPIRE as of 13 May 2014


%\cite{Gavrilov:1996pz}
\bibitem{Gavrilov:1996pz} 
  S.~P.~Gavrilov and D.~M.~Gitman,
  %``Vacuum instability in external fields,''
  Phys.\ Rev.\ D {\bf 53}, 7162 (1996)
  [hep-th/9603152].
  %%CITATION = HEP-TH/9603152;%%
  %74 citations counted in INSPIRE as of 25 Sep 2014

%\cite{Gavrilov:2011ce}
\bibitem{Gavrilov:2011ce} 
  S.~P.~Gavrilov and D.~M.~Gitman,
  %``Creation of neutral fermions with anomalous magnetic moment from vacuum by inhomogeneous magnetic field,''
  arXiv:1101.4243 [hep-th].
  %%CITATION = ARXIV:1101.4243;%%
  %1 citations counted in INSPIRE as of 26 Sep 2014


%\cite{Lin:1998rn}
\bibitem{Lin:1998rn} 
  Q.~-G.~Lin,
  %``Electron - positron pair creation in vacuum by an electromagnetic field in (3+1)-dimensions and lower dimensions,''
  J.\ Phys.\ G {\bf 25}, 17 (1999)
  [hep-th/9810037].
  %%CITATION = HEP-TH/9810037;%%
  %16 citations counted in INSPIRE as of 13 May 2014

%\cite{Lin:1999bb}
\bibitem{Lin:1999bb} 
  Q.~-G.~Lin,
  %``Pair creation of neutral particles in a vacuum by external electromagnetic fields in (2+1)-dimensions,''
  J.\ Phys.\ G {\bf 25}, 1793 (1999)
  [hep-th/9909045].
  %%CITATION = HEP-TH/9909045;%%
  %9 citations counted in INSPIRE as of 13 May 2014

%\cite{Lin:2000rh}
\bibitem{Lin:2000rh} 
  Q.~-G.~Lin,
  %``Vacuum polarization for neutral particles in (2+1)-dimensions,''
  J.\ Phys.\ G {\bf 26}, L17 (2000)
  [hep-th/0002178].
  %%CITATION = HEP-TH/0002178;%%
  %1 citations counted in INSPIRE as of 13 May 2014


%\cite{Chair:2000vb}
\bibitem{Chair:2000vb} 
  N.~Chair and M.~M.~Sheikh-Jabbari,
  %``Pair production by a constant external field in noncommutative QED,''
  Phys.\ Lett.\ B {\bf 504}, 141 (2001)
  [hep-th/0009037].
  %%CITATION = HEP-TH/0009037;%%
  %41 citations counted in INSPIRE as of 06 Jul 2014

%\cite{Brezin:1970xf}
\bibitem{Brezin:1970xf} 
  E.~Brezin and C.~Itzykson,
  %``Pair production in vacuum by an alternating field,''
  Phys.\ Rev.\ D {\bf 2}, 1191 (1970).
  %%CITATION = PHRVA,D2,1191;%%
  %265 citations counted in INSPIRE as of 06 Jul 2014

 
%\cite{Adorno:2014bsa}
\bibitem{Adorno:2014bsa} 
  T.~C.~Adorno, S.~P.~Gavrilov and D.~M.~Gitman,
  %``Particle creation from the vacuum by an exponentially decreasing electric field,''
  arXiv:1409.7742 [hep-th].
  %%CITATION = ARXIV:1409.7742;%%

%\cite{Hounkonnou:2000im}
\bibitem{Hounkonnou:2000im} 
  M.~N.~Hounkonnou and M.~Naciri,
  %``Production of Dirac particles in vacuum by external fields in d + 1 (d = 3,2,1) dimensions,''
  J.\ Phys.\ G {\bf 26}, 1849 (2000).
  %%CITATION = JPHGB,G26,1849;%%
  %6 citations counted in INSPIRE as of 13 May 2014




%\cite{Hounkonnou:1999ym}
\bibitem{Hounkonnou:1999ym} 
  M.~N.~Hounkonnou and J.~E.~B.~Mendy,
  %``Exact solutions of Dirac equation for neutrinos in presence of external fields,''
  J.\ Math.\ Phys.\  {\bf 40}, 4240 (1999).
  %%CITATION = JMAPA,40,4240;%%
  %4 citations counted in INSPIRE as of 13 May 2014

%\cite{BenGeloun:2006uv}
\bibitem{BenGeloun:2006uv} 
  J.~Ben Geloun, J.~Govaerts and M.~N.~Hounkonnou,
 % ``Bosonization of the Schwinger Model by Noncommutative Chiral Bosons,''
  hep-th/0608024.
  %%CITATION = HEP-TH/0608024;%%

%\cite{Michael}
\bibitem{Michael}
M.~Weyrauch, D.~Scholz,
%``Computing the Baker-Campbell-Hausdorff series and the Zassenhaus product,''
J.\ Comput.\ Phys.\ Commun.\ {\bf 180}, 1558-1565 (2009).


%\cite{Casas}
\bibitem{Casas}
F.~Casas, A.~Murua, M.~ Nadinic
%``Efficient computation of the Zassenhaus formula,''
J.\ Comput.\ Phys.\ Commun.\ {\bf 183}, 2386-2391 (2012).
\end{thebibliography}
\end{document}